\documentstyle[12pt]{article}

\begin{document}

\title{Mass relation of $\rho$ and $a_{1}$ mesons}
\author{Bing An Li\\
Department of Physics and Astronomy, University of Kentucky\\
Lexington, KY 40506, USA}

\maketitle

\begin{abstract}

A mass relation of $\rho$ and $a_{1}$ mesons has been obtained
from Weinberg's first sum rule and KSFR sum rule.
\end{abstract}

\newpage
One of the most important features revealed from quantum chromodynamics
(QCD) is the chiral symmetry in the limit of $m_{q}\rightarrow 0$
(\(q=u,d,s\)). On the other hand, current algebra is a successful
theory and Vector Meson Dominance(VMD) is fruitful in studying
electromagnetic properties of hadrons. We abbreviate chiral
symmetry, current algebra, and VMD as CHCV. CHCV are powerful tools
used to study hadron physics at low energies(nonperturbative $QCD$).
Effective chiral theory of mesons[1] is one of important approaches
used to study nonperturbative $QCD$.
CHCV constrain the structure of effective chiral theory of mesons.
$\rho$ and $a_{1}$ mesons have been treated as chiral partners.
Weinberg's first sum rule[2] has been established by CHCV. By invoking
an additional assumption, Weinberg's second sum rule[2]
\begin{equation}
g_{\rho}=g_{a}
\end{equation}
has been found. The mass relation
\begin{equation}
m^{2}_{a}=2m^{2}_{\rho}
\end{equation}
is the consequence of eq.(1) and the KSFR sum rule[3]
\begin{equation}
g^{2}_{\rho}={1\over 2}F^{2}_{\pi}m^{2}_{\rho}.
\end{equation}
The mass formula(eq.(2)) predicts that \(m_{a}=1089MeV\) and the data[4]
shows \(m_{a}=1230\pm 40MeV\). In this letter we try to improve
the mass relation(eq.(2)) based on CHCV only.

By using CHCV, Weinberg's first sum rule[2]
\begin{equation}
\frac{g_{\rho}^{2}}{m_{\rho}^{2}}-\frac{g_{a}^{2}}{m_{a}^{2}}
=\frac{F_{\pi}^{2}}{4}
\end{equation}
has been derived, where
\[<0|V^{a}_{\mu}|\rho^{\lambda}_{b}>=\epsilon^{\lambda}_{\mu}
\delta_{ab}g_{\rho},\]
\[<0|A^{a}_{\mu}|a^{\lambda}_{b}>=\epsilon^{\lambda}_{\mu}
\delta_{ab}g_{a},\]
and $F_{\pi}$ is pion decay constant, \(F_{\pi}=186MeV\).
This sum rule(eq.(4)) can be tested.
$g_{\rho}$ is the coupling constant between
$\rho$ and $\gamma$ and it has been determined to be
\begin{equation}
g_{\rho}=0.116(1\pm 0.05)GeV^{2}
\end{equation}
from $\Gamma(\rho\rightarrow l^{+}l^{-})$.
Using \(m_{a}=(1230\pm 40)MeV\), the eq.(4) predicts that
\begin{equation}
g_{a}=0.145\pm 0.018 GeV^{2}.
\end{equation}
{}From eqs.(5,6) it can be seen that Weinberg's second sum rule(eq.(1))
is not in good agreement with data.
On the other hand, $g_{a}$ can be determined from
\begin{equation}
\Gamma(\tau\rightarrow a_{1}\nu)=\frac{G^{2}}{8\pi}cos^{2}
\theta_{c}g^{2}_{a}\frac{m^{3}_{\tau}}{m^{2}_{a}}(1-
\frac{m^{2}_{a}}{m^{2}_{\tau}})^{2}(1+2\frac{m^{2}_{a}}
{m^{2}_{\tau}}).
\end{equation}
The experimental data of the decay rate is[4] $2.14\times 10^{-13}
(1\pm 0.32)GeV$ and $g_{a}$ is determined to be
$0.145\pm 0.033 GeV^{2}$.
Therefore, Weinberg's first sum rule is in good agreement
with the data. On the
other hand, the KSFR sum rule(eq.(3)) agrees with data within $10\%$.
The combination of eqs.(3,4) leads to a new mass relation between $\rho$
and $a_{1}$. In order to see that let's rewrite eq.(4) as
\begin{equation}
1-\frac{g_{a}^{2}}{g^{2}_{\rho}}\frac{m^{2}_{\rho}}{m_{a}^{2}}
=\frac{F_{\pi}^{2}}{4}\frac{m^{2}_{\rho}}{g^{2}_{\rho}}
\end{equation}
Substituting KSFR sum rule into eq.(8) we obtain
\begin{equation}
m^{2}_{a}=2\frac{g^{2}_{a}}{g^{2}_{\rho}}m^{2}_{\rho}
\end{equation}
Comparing with eq.(2) there is an additional factor of ${g^{2}_{a}\over
g^{2}_{\rho}}$ in eq.(9)
and there is no any additional assumption.
Using the values of $g_{a}$ and $g_{\rho}$(eqs.(5,6)) we
obtain
\begin{equation}
m_{a}= 1.36(1\pm 0.17) GeV
\end{equation}
Theoretical prediction of eq.(9) fits the data of $m_{a}$ better.

To conclude, a new mass relation between $\rho$ and $a_{1}$ mesons
has been presented in terms of Weinberg's first sum rule and KSFR sum
rule.

This research is partially supported by DE-91ER75661.

\end{document}